\documentclass[conference]{IEEEtran}
\IEEEoverridecommandlockouts
\usepackage{cite}
\usepackage{amsmath,amssymb,amsfonts}
\usepackage{algorithmic}
\usepackage{graphicx}
\usepackage{textcomp}
\usepackage{xcolor}
\def\BibTeX{{\rm B\kern-.05em{\sc i\kern-.025em b}\kern-.08em
    T\kern-.1667em\lower.7ex\hbox{E}\kern-.125emX}}
\begin{document}

\title{A distributed system perspective on Backscatter systems\\
{\footnotesize \textsuperscript{*}Note: Sub-titles are not captured in Xplore and
should not be used}

}

\author{\IEEEauthorblockN{1\textsuperscript{st} Tonghuan Xiao}
\IEEEauthorblockA{\textit{Institute of Advanced Technology} \\
\textit{University of Science and Technology of China}\\
AnHui, China \\
1828748526@qq.com}
\and
\IEEEauthorblockN{2\textsuperscript{nd} Jiecheng Zhou}
\IEEEauthorblockA{\textit{School of Information Science and Technology} \\
\textit{University of Science and Technology of China}\\
Anhui, China \\
zhoujiecheng@mail.ustc.edu.cn}

}

\maketitle

\begin{abstract}
This review investigates the pivotal role of distributed architectures and intelligent resource allocation in enabling robust and scalable wireless systems, with a particular emphasis on backscatter communication, indoor localization, battery-free networks, and Simultaneous Wireless Information and Power Transfer (SWIPT). Recent advancements—such as Relacks, HitchHike, and Inter-Technology Backscatter—demonstrate how distributed components and commodity hardware can be leveraged for innovative and dependable backscatter applications. In localization, B2Loc illustrates the power of collaborative sensing with distributed BLE infrastructures. For battery-free networks, protocols like Find and environmental synchronization mechanisms such as Flync address the challenges of intermittent operation. The review also delves into resource allocation for multi-user SWIPT, highlighting the profound impact of realistic energy harvester models on system design. Furthermore, the reliability of backscatter networks is examined as a critical enabler for large-scale, mission-critical IoT applications. Representative approaches—including macro-level system co-design, collision detection and avoidance, coding strategies, and physical-layer innovations—are discussed as means to enhance communication integrity and operational resilience. This synthesis provides a comprehensive overview of current trends and future directions for distributed, reliable, and energy-efficient wireless networks.

\end{abstract}

\begin{IEEEkeywords}
Distributed Architectures,
Resource Allocation,
Backscatter Communication,
Wireless Information and Power Transfer (WIPT),
Simultaneous Wireless Information and Power Transfer (SWIPT),
Commodity Hardware,
\end{IEEEkeywords}

\section{Introduction}

The rapid expansion of the Internet of Things (IoT) has ushered in an era where billions of devices are interconnected, driving demand for innovative communication paradigms that are both energy-efficient and scalable. Low-power communication techniques, particularly backscatter communication and wireless power transfer, have emerged as pivotal technologies to meet these demands, enabling devices to operate with minimal or even no batteries by harvesting ambient energy or reflecting existing radio frequency (RF) signals \cite{b8, b13}. However, the effective deployment of these technologies, especially in scenarios involving a massive number of resource-constrained devices operating in dynamic and often unpredictable environments, presents significant challenges for traditional centralized network models. Such models often struggle with scalability, robustness, and efficient resource utilization.

This review delves into the critical roles of \textbf{distributed architectures} and \textbf{intelligent resource allocation strategies} in overcoming these challenges and unlocking the full potential of emerging wireless systems. We explore a range of approaches presented in contemporary literature, focusing on how distributed designs and sophisticated resource management can enhance the performance, reliability, and practicality of these systems. The scope of this review covers several key areas:
\begin{itemize}
    \item \textbf{Reliable backscatter communication}, where coordinating distributed transceivers and tags is essential for maintaining robust links in complex indoor environments \cite{b8}.
    \item \textbf{Accurate indoor localization} using Bluetooth Low Energy (BLE) backscatter, which leverages distributed BLE infrastructures to achieve precise positioning of low-power tags \cite{b11}.
    \item \textbf{Battery-free wireless networking}, focusing on protocols and synchronization mechanisms that enable device-to-device communication among intermittently powered nodes \cite{b12}.
    \item \textbf{Simultaneous Wireless Information and Power Transfer (SWIPT)}, particularly in multi-user scenarios where distributed beamforming and interference management are crucial for balancing the dual objectives of data transmission and energy delivery \cite{b13}.
\end{itemize}
Reliability stands out as a central requirement for the practical application of these technologies, directly impacting data integrity, system responsiveness, and safety in mission-critical IoT scenarios such as environmental sensing, asset tracking, and healthcare monitoring. This review also highlights recent research on improving reliability from multiple perspectives: macro-level system co-design, advanced collision detection and avoidance, coding and signal processing techniques, and physical-layer innovations. By synthesizing these advances, we present an integrated perspective on how distributed architectures and intelligent resource allocation can achieve both high efficiency and high reliability, paving the way for next-generation wireless networks.

\section*{Distributed Architecture and Resource Allocation}

The proliferation of Internet of Things (IoT) devices, coupled with advancements in low-power communication techniques such as backscatter and wireless power transfer, necessitates a paradigm shift towards more sophisticated distributed architectures and intelligent resource allocation strategies. Traditional centralized network models often fall short in terms of scalability, robustness, and efficiency when dealing with massive numbers of resource-constrained devices operating in dynamic and often unpredictable environments. This review explores various approaches to distributed architectures and resource allocation as presented in the provided literature, covering areas from reliable backscatter communication and accurate indoor localization to battery-free networking and simultaneous wireless information and power transfer (SWIPT).

\subsection*{Distributed Architectures in Backscatter Communication}

Backscatter communication, by its nature, often involves distributed elements, especially in bistatic configurations where the radio frequency (RF) source and the receiver are separate entities. The challenge in such systems is to coordinate these elements effectively to ensure reliable communication.

\textbf{Relacks: A Closed-Loop Bistatic Backscatter System}

\begin{figure}
    \centering
    \includegraphics[width=1\linewidth]{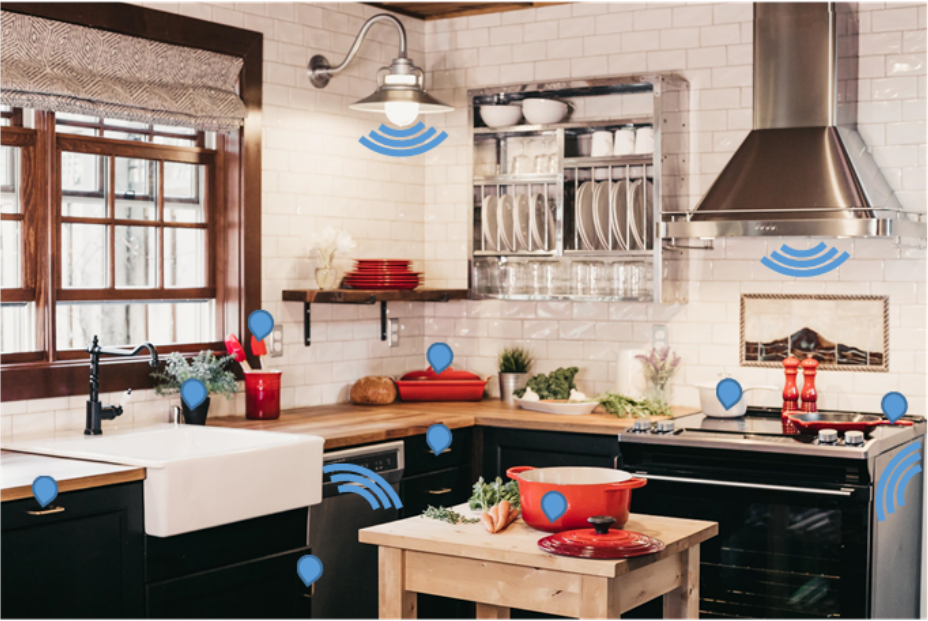}
    \caption{indoor communication}
    \label{fig:fig-1-1}
\end{figure}

\begin{figure}
    \centering
    \includegraphics[width=1\linewidth]{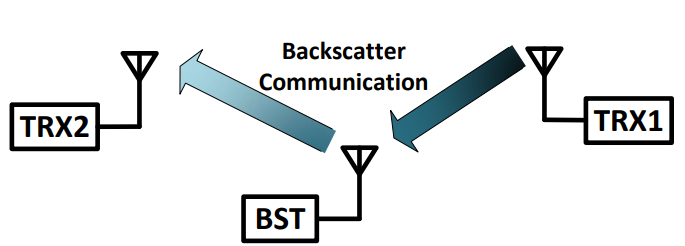}
    \caption{The TRX boards follow the defined settings to communicate with the tag.}
    \label{fig:fig-1-2}
\end{figure}

\begin{figure}
    \centering
    \includegraphics[width=1\linewidth]{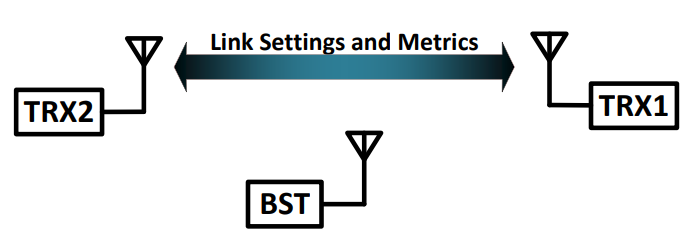}
    \caption{The TRX boards share link settings and metrics to avoid using low quality links for future backscatter communications.}
    \label{fig:fig-1-3}
\end{figure}

The Relacks system, as detailed in Katanbaf et al., proposes a closed-loop bistatic backscatter architecture designed for reliable indoor(shown in Fig.~\ref{fig:fig-1-1}) communication \cite{b8} . This system inherently employs a distributed architecture consisting of two transceiver (TRX) base units and one or more backscatter tags. A key challenge in bistatic backscatter is the \textit{distributed nature of link quality assessment}; the TRX providing the excitation signal cannot unilaterally determine if a backscatter packet has been successfully formed and received by the other TRX. Relacks addresses this by having the TRX boards share link settings and metrics. This shared information is crucial for avoiding low-quality links in future communications, thereby improving overall reliability (shown in Fig.~\ref{fig:fig-1-3} and Fig.~\ref{fig:fig-1-2}).

\begin{figure}
    \centering
    \includegraphics[width=1\linewidth]{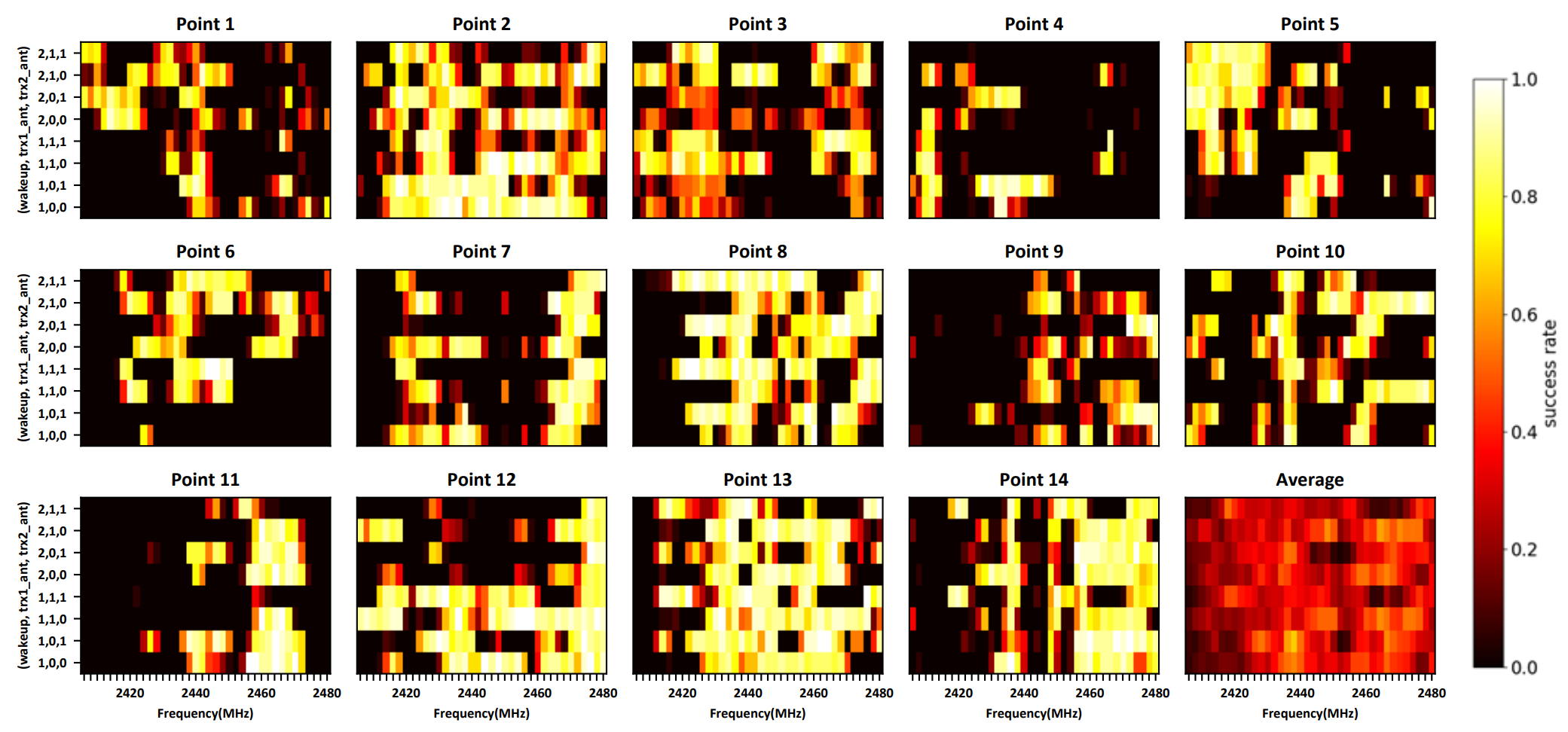}      
    \caption{Success rate variation by using different configurations}
    \label{fig:fig-1-4}
\end{figure}

Resource allocation in Relacks is managed by a central controller connected to one of the TRXs (designated as the receiver). This controller makes decisions based on the history of previous communications, selecting parameters such as communication frequency, transceiver antennas, and the wake-up source. The paper highlights three main diversity sources exploited for resource allocation:
\begin{itemize}
    \item \textbf{Frequency Diversity}: Exploiting the 80 MHz bandwidth of the 2.4 GHz ISM band. The system can choose from 38 frequency options (assuming 2MHz spacing and a 3.5 MHz separation between RF source and receiver frequencies). The success rate heatmaps(Fig.~\ref{fig:fig-1-4})  demonstrate significant channel variation across frequencies.
    \item \textbf{Antenna Diversity}: Each TRX board is equipped with two omnidirectional dipole antennas. This provides four antenna configuration options, which helps mitigate spatial-selective fading. The importance of antenna diversity is particularly noted for specific points where one antenna shows poor link quality while the other provides high success rates.
    \item \textbf{Wake-up Source Diversity}: One of the two TRXs is selected to wake up the tag. This is crucial as simultaneous wake-up transmissions would cause interference. Ideally, the closer TRX should perform the wake-up, but tag location is not always known.
\end{itemize}
The proposed algorithm dynamically selects these communication parameters, achieving a significant improvement in success rate compared to random selection. The system aims for uniform coverage over areas up to 60 $m^2$, even in worst-case arrangements where the tag is not directly between the TRXs. This distributed coordination and adaptive resource allocation are fundamental to achieving reliable backscatter communication in multipath-rich indoor environments.

\textbf{HitchHike: Backscatter Using Commodity WiFi Infrastructure}

Zhang et al. in "HitchHike: Practical Backscatter Using Commodity WiFi" present a system that leverages existing WiFi infrastructure, inherently a distributed architecture\cite{b9}. In a typical HitchHike deployment, an excitation device (e.g., a smartphone) transmits a standard 802.11b packet on one WiFi channel (e.g., Channel 1) to a primary Access Point (AP1). A low-power HitchHike tag reflects this signal, modulating its own information, and frequency shifts it to an adjacent non-overlapping WiFi channel (e.g., Channel 6). A secondary AP (AP2), tuned to this adjacent channel, receives and decodes the backscattered packet.
The resource allocation in this scenario involves the selection of WiFi channels for the primary transmission and the backscattered signal to avoid interference and ensure spectral efficiency. The system's ability to utilize existing, geographically distributed APs highlights its distributed nature. The decoding process, particularly the XOR decoding to retrieve the tag's data, can also be distributed: AP2 sends the decoded backscattered packet to AP1, which then performs the XOR operation with the original packet. This coordination between distributed APs is crucial for the system's operation. Furthermore, HitchHike's design for co-existence with existing WiFi networks is a form of implicit resource sharing in the unlicensed spectrum.

\textbf{Inter-Technology Backscatter}

\begin{figure}
    \centering
    \includegraphics[width=1\linewidth]{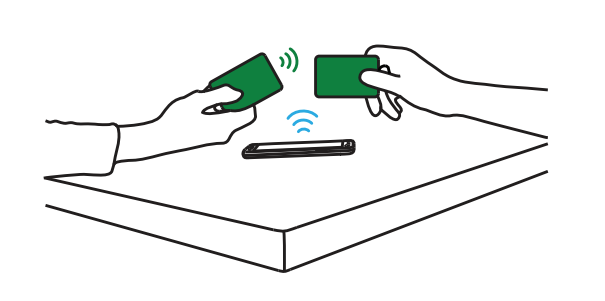}  
    \caption{device-to-device nature of Inter-Technology Backscatter}
    \label{fig:fig-1-5}
\end{figure}

Iyer et al. introduce "Inter-Technology Backscatter\cite{b10}," which transforms wireless transmissions from one technology (e.g., Bluetooth) to another (e.g., Wi-Fi or ZigBee) on the air. This system leverages commodity devices like smartphones and smartwatches as distributed components for both RF signal generation and reception. For example, a smartwatch transmitting Bluetooth signals can act as the RF source, while a smartphone receives the Wi-Fi signals generated by the backscatter device. This approach allows for opportunistic use of ubiquitously available devices, forming a dynamically structured distributed system. Resource management here involves selecting appropriate source (e.g., Bluetooth advertising channels) and target (e.g., Wi-Fi channels) frequencies to ensure compatibility and minimize interference. Applications such as card-to-card communication, enabled by backscattering Bluetooth transmissions from a nearby smartphone, further underscore the distributed, device-to-device nature of this architecture (shown in Fig.~\ref{fig:fig-1-5}).

\subsection*{Distributed Architectures for Indoor Localization}

Accurate indoor localization often relies on signals from multiple, spatially distributed reference points. Recent advancements in BLE backscatter technology are enabling low-power localization solutions that leverage existing or easily deployable BLE infrastructures.

\textbf{B\textsuperscript{2}Loc: BLE Backscatter Localization}

Luo et al. propose B\textsuperscript{2}Loc, a system for decimeter-level indoor localization of low-power BLE backscatter tags using existing BLE infrastructures\cite{b11}. The architecture is inherently distributed, relying on a group of ubiquitous BLE receivers (e.g., beacons, smartphones) to collaboratively localize a tag. Each BLE receiver, equipped with an antenna array, estimates the Angle-of-Arrival (AoA) of the backscattered signals.
Resource allocation and coordination involve:
\begin{itemize}
    \item \textbf{AoA Estimation and Sanitization}: Each distributed receiver independently estimates AoA. B\textsuperscript{2}Loc employs a time-frequency cooperative sanitization approach to mitigate multipath effects by leveraging the observation that AoA peaks from reflections are more sensitive to time variations and channel shifts than the direct path.
    \item \textbf{Confidence-Aware Localization}: The system combines AoA estimates from multiple infrastructures. A confidence-aware localization algorithm assigns weights to different AoA peaks based on their diffusion, favoring sharper, more concentrated peaks that are more likely to correspond to the direct path and intersect at the true tag position.
    \item \textbf{Transceiver-Reversal}: To improve localization, especially when the downlink from a specific transmitter to the tag is weak, B\textsuperscript{2}Loc allows infrastructures to alternate roles. Any infrastructure can act as an excitation source, while others act as receivers. This dynamic reallocation of roles enhances the probability of obtaining strong signal paths.
    \item \textbf{Impact of Receiver Density}: The density and placement of BLE receivers (a distributed resource) significantly impact localization accuracy. Denser deployments increase the likelihood of capturing strong direct paths and improve triangulation. Experiments show that increasing the number of receivers from two to four improves median localization error.
\end{itemize}
The system architecture involves deploying multiple BLE beacons as receivers, demonstrating a practical distributed setup.

\subsection*{Resource Allocation in Battery-Free Wireless Networks}

Battery-free devices, operating intermittently based on harvested energy, present unique challenges for network formation and communication. Distributed protocols and synchronization mechanisms are crucial for these devices to discover each other and coordinate their activities.

\textbf{Find and Flync: Neighbor Discovery and Synchronization}

\begin{figure}
    \centering
    \includegraphics[width=1\linewidth]{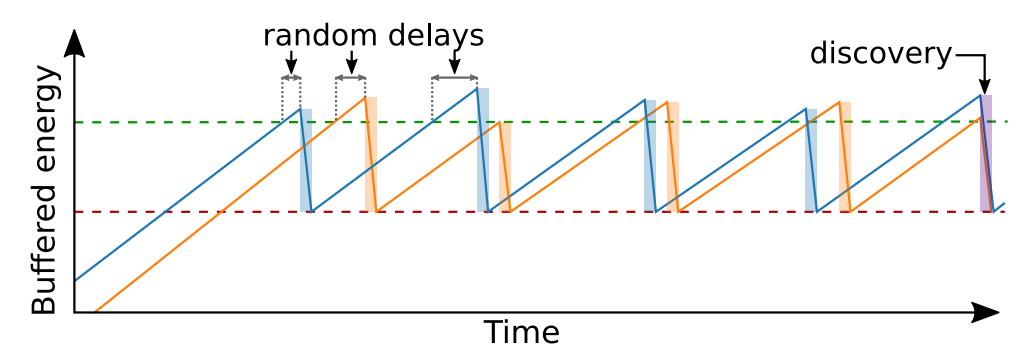}
    \caption{Using Find, nodes randomly delay their wake-ups to avoid interleaving, thereby discovering each other faster and more efficiently.}
    \label{fig:fig-1-6}
\end{figure}

Geissdoerfer and Zimmerling address the problem of efficient device-to-device communication in intermittently powered battery-free wireless networks\cite{b12}. They introduce \textbf{Find}, a neighbor discovery protocol, and \textbf{Flync}, a hardware/software solution for synchronization.
The core challenge is that battery-free nodes have interleaved active phases due to the need to charge their capacitors.
\begin{itemize}
    \item \textbf{Find Protocol}: This protocol uses randomized waiting to minimize discovery latency. Instead of becoming active immediately upon reaching the turn-on energy threshold (greedy approach), nodes delay their wake-up by a random time drawn from an optimized geometric distribution(shown in Fig.~\ref{fig:fig-1-6}). This desynchronizes their wake-up patterns. The scale parameter of this distribution is dynamically adapted at runtime by each node based on its own charging time, assuming its neighbors have similar energy availability (a decentralized resource management approach). The modeling of discovery latency for a network of N nodes informs this optimization. The authors show that optimizing for a two-node network often yields good performance across various network densities.
    \item \textbf{Flync Synchronization}: Flync exploits powerline-induced flicker from indoor lamps as an external, common synchronization signal for solar energy harvesting nodes. The Flync circuit extracts a clock signal from the solar panel current variations. When used with Find, nodes implicitly align their wake-ups to this signal, increasing the probability of simultaneous activity and thus speeding up discovery. This is a form of distributed synchronization using a shared environmental resource. The paper evaluates Flync's robustness to mobility, shadowing, and electrical loads, and its performance with different types and arrangements of light sources.
\end{itemize}
The prototype implementation uses custom-designed battery-free nodes and demonstrates significant reductions in discovery latency compared to baseline approaches. These mechanisms are crucial for bootstrapping communication in networks of distributed, intermittently powered devices.

\subsection*{Distributed Architectures and Resource Allocation in Wireless Information and Power Transfer (WIPT)}

The unification of wireless information and power transmission (WIPT) introduces complex resource allocation problems, especially in multi-user scenarios where interference can be both detrimental to information transfer and beneficial for energy harvesting\cite{b13}. Clerckx et al. provide a comprehensive overview of WIPT fundamentals, with a strong emphasis on how energy harvester (EH) models (linear vs. nonlinear) dictate signal and system design. Section IV of their paper is particularly relevant to distributed architectures and resource allocation in multi-user WIPT.

\textbf{Multi-User SWIPT Architectures and Challenges}

WIPT systems can be categorized into SWIPT, Wirelessly Powered Communication Networks (WPCN), and Wirelessly Powered Backscatter Communication (WPBC). Multi-user SWIPT involves a transmitter broadcasting to multiple Information Receivers (IRs) and/or Energy Receivers (ERs), which may be co-located or separated. The key challenge in such distributed systems is managing co-channel interference and allocating resources (power, beams, time, frequency) to satisfy diverse user requirements for rate and energy.

\textbf{Resource Allocation with the Linear EH Model}

Under the conventional linear EH model ($P_{dc}^r = e_3 P_{rf}^r$, where $e_3$ is constant), the transmitted signal $\mathbf{x}$ in a multi-user MISO system can be a superposition of K information beams and J energy beams: $$\mathbf{x} = \sum_{k=1}^{K} \mathbf{p}_k s_k^{ID} + \sum_{j=1}^{J} \mathbf{v}_j s_j^{EH}$$.
\begin{itemize}
    \item \textbf{Beamforming and Interference Management}: The information beams $\mathbf{p}_k$ and energy beams $\mathbf{v}_j$ must be carefully designed. It was shown that if energy signals cannot be cancelled by IRs, no dedicated energy signals should be used ($\mathbf{v}_j = \mathbf{0}$) for optimal R-E tradeoff. The SINR at IR k is then $$\gamma_k = \frac{\mathbf{p}_k^H \mathbf{H}_k \mathbf{p}_k}{\sum_{i \neq k} \mathbf{p}_i^H \mathbf{H}_k \mathbf{p}_i + \sigma^2}$$. Harvested power at ER j is proportional to $\sum_{k=1}^{K} \| \mathbf{g}_j \mathbf{p}_k \|^2$.
    \item \textbf{Optimization Problems}: A common resource allocation problem is to maximize sum-energy harvested by ERs subject to transmit power constraints and SINR constraints at IRs:
    $\max_{\{\mathbf{p}_k\}} \sum_{j=1}^{J} \sum_{k=1}^{K} \| \mathbf{g}_j \mathbf{p}_k \|^2$ subject to $\sum_{k=1}^{K} \|\mathbf{p}_k\|^2 \le P$ and $\gamma_k \ge \bar{\Gamma}_k$. Solutions can be found using Semidefinite Relaxation (SDR) or uplink-downlink duality. Zero-Forcing Beamforming (ZFBF) can serve as an initial, practical approach.
    \item \textbf{Receiver Architectures (TS/PS)}: In multi-user OFDM systems with the linear EH model, Power Splitting (PS) generally outperforms Time Switching (TS). The PS ratios and power allocation across subchannels can be jointly optimized.
\end{itemize}
The main observation for the linear model in multi-user SWIPT is that strategies maximizing total received RF power also maximize total harvested DC power, and CSCG inputs are typically optimal.

\textbf{Resource Allocation with Nonlinear EH Models}

When more realistic nonlinear EH models (diode nonlinear model or saturation nonlinear model) are considered, resource allocation strategies change significantly.
\begin{itemize}
    \item \textbf{Saturation Nonlinear Model}: The harvested power $P_{dc,j}^r$ at ER j is a nonlinear (e.g., sigmoidal) function of its received RF power $P_{rf,j}^r = \sum_{k=1}^{K} \text{Tr}(\mathbf{p}_k \mathbf{p}_k^H \mathbf{g}_j^H \mathbf{g}_j)$. The sum-power maximization problem becomes $\max_{\{\mathbf{p}_k\}} \sum_{j=1}^{J} P_{dc,j}^r$. This problem is more complex due to the sum-of-ratios form (if the sigmoid is expanded). Iterative algorithms can solve it by transforming the objective function.
    The optimal beamforming structure differs from the linear model. For instance, even with orthogonal ER channels, the saturation model may not allocate all power to the strongest ER if it's already saturated. Instead, it might transmit via an eigenmode of a weighted sum of channel matrices, where weights (dual variables) decrease for ERs nearing saturation.
    \item \textbf{Diode Nonlinear Model}: While multi-user SWIPT for the diode nonlinear model is less explored, insights from single-user cases suggest significant impacts. The strategy maximizing $P_{rf}^r$ will \textit{not} maximize $P_{dc}^r$. Optimal input distributions will likely deviate from CSCG, potentially favoring asymmetric or non-zero mean inputs to exploit the diode's nonlinearity and boost higher-order moments of the received signal, as discussed for single-user scenarios. This implies that waveform design, modulation, and power allocation across users and subcarriers will need to be fundamentally rethought. Interference, which contains power, could also be shaped to be more "harvestable" under this model.
\end{itemize}
A key takeaway is that for the saturation model, maximizing total received RF power does \textit{not} maximize total harvested DC power, unlike the linear model. The beamforming direction is generally different. For the diode nonlinear model, the beneficial effects of nonlinearity seen in single-user cases are expected to extend to multi-user scenarios, leading to potentially larger R-E regions if resources are allocated optimally.

\textbf{Future Directions in Multi-User WIPT Resource Allocation}
Clerckx et al. highlight several open research areas for multi-user WIPT with nonlinear EHs:
\begin{itemize}
    \item Developing multi-user SWIPT waveform, modulation, and input distribution designs for the diode nonlinear model.
    \item Investigating fundamental limits (e.g., capacity regions) of multi-user channels under nonlinear EH models.
    \item Re-evaluating SWIPT architectures (broadcast, multiple access, interference, relay channels) with nonlinearities.
    \item Studying the impact of EH nonlinearities on WPCN and WPBC in multi-user settings. For example, multi-user WPBC waveform design needs to maximize SINR at the reader and harvested energy at tags, considering diode nonlinearity, channel diversity, and multi-user diversity.
    \item Exploring advanced interference management techniques like \textbf{rate-splitting multiple access (RSMA)} for multi-user SWIPT under both linear and nonlinear EH models. RSMA could offer a more flexible framework to manage interference for both information decoding and energy harvesting.
\end{itemize}

\subsection*{Synchronization and Coordination in Distributed Systems}
Effective operation of distributed wireless systems heavily relies on synchronization and coordination among nodes. This is particularly challenging in energy-constrained and intermittent networks.

\textbf{Synchronization in Relacks}
The Relacks system requires time synchronization between the two TRX boards and the tag for successful backscatter communication. Before the backscatter phase, an active BLE packet is sent from the receiver TRX to the transmitter TRX. This packet shares configuration information (frequency, antenna, wake-up source, tag ID) and serves as a primary synchronization mechanism. A subsequent wake-up packet from one TRX to the tag signals the tag to exit low-power mode and further aids synchronization. This entire process, including guard intervals, is repeated before each backscatter communication phase to maintain robust synchronization in a dynamic environment. The overhead of this synchronization depends on the number of backscatter packets transmitted per wake-up.

\textbf{Synchronization in HitchHike}
The HitchHike tag needs to synchronize its backscattering operation with the incoming 802.11b packets. It uses an analog envelope detector to identify the start of an 802.11b transmission from the commodity WiFi transmitter. To avoid corrupting the 802.11b preamble and header, the tag waits for a deterministic period after detecting the start of an excitation packet before it begins backscattering its data. The system also employs a synchronization phase where the 802.11b transmitter sends a sequence of short packets, which the tag decodes using OOK via its envelope detector, to know when to expect the main excitation packet. The timing precision of this envelope detector is critical; experiments showed that the introduced delay is typically less than 2$\mu$s, which does not significantly impact decoding performance.

\textbf{Flync for External Synchronization}
As discussed earlier, Flync provides a novel way for distributed solar-harvesting nodes to achieve phase synchronization by using the powerline-induced flicker of ambient lights. This external, globally available signal (within a building or area served by the same power phase) allows nodes to implicitly align their activity periods without direct communication, which is highly beneficial for energy-constrained, intermittent devices. The use of a PLL in software to lock onto the extracted flicker signal ensures stability despite disturbances like MPPT cycles or temporary shadowing. This distributed synchronization primitive significantly accelerates neighbor discovery when combined with protocols like Find. However, the phase of this flicker can vary if lights are connected to different phases of a three-phase power system, which needs to be accounted for if nodes are in different rooms.

\section{Reliability in backscatter networks}

Reliability is a cornerstone for the practical deployment and scalability of 
distributed system and backscatter networks, as it directly affects the integrity and utility of data transmission in diverse Internet of Things (IoT) applications. In critical use cases such as environmental sensing, asset tracking, and healthcare monitoring, unreliable communication can result in data loss, misinterpretation, and system failures, potentially leading to safety hazards or significant operational inefficiencies \cite{b14, b15}. The inherently low-power and passive nature of backscatter devices, which communicate by reflecting ambient radio frequency signals rather than generating their own, makes them more vulnerable to signal attenuation, interference, and environmental dynamics, thereby exacerbating the risk of communication errors \cite{b16}. High reliability ensures that such networks can deliver consistent and accurate information, supporting real-time decision-making and automation in large-scale deployments, especially where devices are often unattended and exposed to unpredictable conditions \cite{b14, b17}. Furthermore, reliable backscatter communication is essential for seamless integration with existing wireless infrastructures and for achieving the energy efficiency necessary for long-term, maintenance-free operation \cite{b15}. Thus, enhancing reliability is not only crucial for maintaining data fidelity and system responsiveness but also for unlocking the full potential of backscatter networks in mission-critical and resource-constrained scenarios. A substantial body of literature has contributed to enhancing the reliability of backscatter networks. In the following, we will discuss several representative works, each of which improves network reliability from different perspectives. Firstly ,we discuss the reliability in macro-level system design, and then, we list some optimizations in enhancing reliability from the perspective of collision detection and avoidance, finally, we discuss measures to enhance reliability from the perspectives of coding and physical materials.

\begin{figure*}[t]
    \centering
    \includegraphics[height=20\baselineskip]{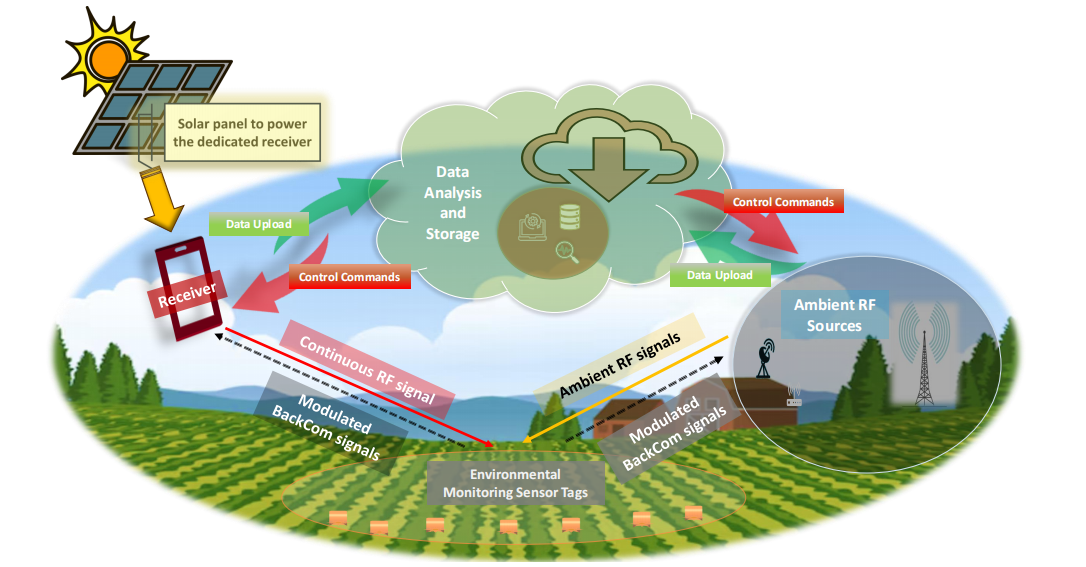}
    \caption{An Example of Hybrid Backscatter Communication System.}
    \label{fig0g}
\end{figure*}

\cite{b18} emphasizes several key technologies from a macro-level system design perspective to enhance the reliability of BackCom-enabled 0G networks. As shown in figure \ref{fig0g}, first, it advocates for \textbf{hybrid backscatter architectures} that combine dedicated and ambient RF sources (e.g., Wi-Fi, LoRa, LTE), ensuring continuous and stable communication even when some sources are unreliable. Advanced modulation schemes, such as Frequency-Shift Keying (FSK), Phase-Shift Keying (PSK), spatial modulation, and adaptive modulation techniques, are highlighted to improve resilience against interference and multipath fading, thus lowering error rates. The integration of multi-antenna systems and beamforming further strengthens signal robustness, especially in noisy or dynamic environments. The 0G network’s wide coverage and \textbf{centralized management} provide consistent RF power and reliable data transmission, reducing communication outages due to environmental or topological changes. The introduction of the Waste Factor (WF) metric enables systematic energy optimization, indirectly supporting reliability by minimizing power-related disruptions. In practical deployments, \textbf{redundant sensor placement and real-time monitoring} are recommended to ensure data continuity in harsh or remote conditions. The paper also points to future directions such as AI-driven adaptive protocols and cognitive spectrum management, which can dynamically optimize network parameters in response to changing conditions, further bolstering the overall reliability of large-scale, green IoT networks.

\begin{figure}[htbp]
\centerline{\includegraphics[height=8\baselineskip]{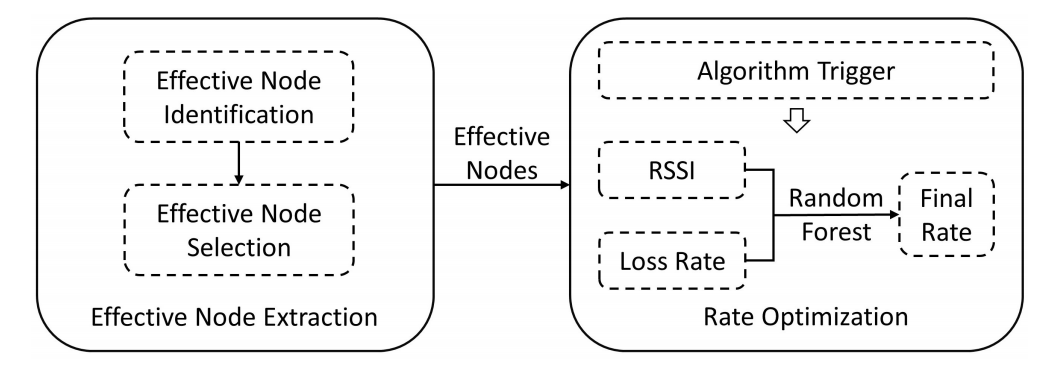}}
\caption{The overview of RAEN system.}
\label{figraen}
\end{figure}

\begin{figure}[htbp]
\centerline{\includegraphics[height=12\baselineskip]{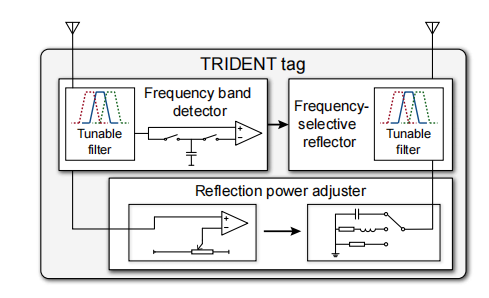}}
\caption{The overview of Trident system.}
\label{fig_trident}
\end{figure}

    \cite{b19} proposes RAEN, a reliability-oriented rate adaptation algorithm specifically designed for effective nodes in large-scale backscatter networks. As illustrated in the figure \ref{figraen}, the core of RAEN is its two-stage approach: first, it accurately identifies and selects effective nodes—those with valuable or dynamic information—by modeling each node’s RF phase behavior using a renewable Gaussian model. This model allows the system to detect environmental changes or node mobility, ensuring that only nodes with time-varying or critical data are prioritized. Second, RAEN employs a novel trigger mechanism based on inventory cycles, reducing protocol overhead by eliminating the need for constant channel probing. For rate selection, a random forest classifier leverages real-time indicators like RSSI and packet loss rate to dynamically predict and assign the optimal transmission rate for the group of effective nodes. This strategy not only increases individual throughput but also avoids network congestion and collisions that degrade reliability in dense environments. Extensive experiments using commercial RFID equipment show that RAEN can triple effective node throughput compared to traditional methods, all while remaining fully compatible with EPC C1G2 standards. Thus, RAEN significantly enhances the reliability and efficiency of backscatter network communications in dynamic and large-scale IoT scenarios. \cite{b22} introduces Trident, a novel reliability technology for multi-reader backscatter networks that significantly enhances interference avoidance and network throughput. Unlike traditional TDMA or CSMA approaches, which separate reader access only in the time domain and suffer from low concurrency and throughput, Trident leverages frequency-space division to enable concurrent, interference-free operation of multiple readers and tags. As presented in figure \ref{fig_trident} Trident tags are equipped with a frequency band detector and a frequency-selective reflector, enabling them to detect the strongest excitation frequency band and selectively backscatter signals only in that band. Additionally, a reflection power adjuster dynamically controls the strength of the backscattered signal, preventing excessive interference to distant readers operating at the same frequency. The system also introduces a frequency assignment algorithm based on an interference graph and genetic optimization, ensuring adjacent readers operate on different bands and minimizing co-channel interference even in dense or irregular deployments. Experimental results show Trident can boost network throughput by up to 3.18× compared to TDMA, while maintaining low power consumption suitable for battery-free tags. Overall, Trident’s combination of adaptive frequency selection, power control, and intelligent frequency allocation provides a robust and scalable solution for reliable, high-density backscatter communications.

\cite{b20} focusing on interference suppression and detection optimization at the signal level to address reliability bottlenecks at the link layer, presents a novel interference cancellation scheme to significantly improve the reliability of bistatic backscatter communication systems. The main challenge addressed is the strong direct-link interference between the carrier emitter and the reader, which often overwhelms the weak backscattered signal from the tag and degrades detection performance. The authors propose \textbf{a joint design of the carrier and tag coding structure}, exploiting the periodicity of the RF carrier and using Miller coding at the tag. This enables the receiver to perform simple algebraic manipulation on the sampled signals, effectively cancelling out the direct-link interference without relying on complex hardware or carrier synchronization, and thus mitigating issues such as carrier frequency offset. Based on this interference-free signal, the paper develops an optimal detector with a closed-form bit error rate (BER) expression, which consistently outperforms the conventional energy detector benchmark in both theoretical analysis and simulations. An efficient parameter estimation method is also introduced to ensure practical implementation. Simulation results confirm that the proposed scheme provides robust and reliable detection even under imperfect parameter estimation and in the presence of channel fading, thus greatly enhancing the reliability of bistatic backscatter systems for practical IoT applications. This paper 

\cite{b21} investigates the enhancement of reliability for orbital angular momentum (OAM) vortex beams generated by ultra-large-scale metasurfaces, focusing on their anti-interference capabilities. The core technology lies in increasing the physical size of the metasurface, which, according to theoretical modeling and full-wave simulations, leads to stronger resistance against noise and interference. As the metasurface becomes larger, the resulting OAM beams exhibit more concentrated energy distribution, fewer sidelobes, and higher mode purity in both amplitude and phase, even in the presence of significant Gaussian noise. The study demonstrates that small metasurfaces are susceptible to noise, with vortex field characteristics quickly degraded, while ultra-large metasurfaces maintain their OAM properties and spectrum purity under identical noise conditions. The paper also details the design and simulation of high-reflectivity, broadband EE unit cells that constitute the metasurface, ensuring efficient generation and manipulation of OAM beams. These findings suggest that scaling up metasurface dimensions is a practical method to achieve robust, high-purity OAM beam transmission, which is critical for reliable high-capacity communication systems. The technology and simulation methodologies presented provide valuable insights for improving the \textbf{physical-layer reliability} of advanced wireless and backscatter networks.

\section*{Conclusion}

This review has traversed a diverse landscape of emerging wireless communication systems, underscoring the pivotal role of \textbf{distributed architectures} and \textbf{intelligent resource allocation} in enabling their practical deployment and optimal performance. From enhancing the reliability of bistatic backscatter systems like Relacks through coordinated transceiver operation and dynamic parameter selection \cite{b8}, to leveraging ubiquitous commodity WiFi and Bluetooth infrastructures for novel backscatter communication in HitchHike and Inter-Technology Backscatter \cite{b9, b10}, the trend towards decentralized and opportunistic system design is evident. Similarly, in the realm of indoor localization, systems such as B\textsuperscript{2}Loc demonstrate that accurate positioning can be achieved by harnessing the collective intelligence of distributed BLE receivers, employing sophisticated AoA estimation, sanitization, and confidence-aware data fusion techniques \cite{b11}.

The unique challenges posed by battery-free networks, characterized by intermittent operation, are effectively addressed by distributed protocols like Find, which employs randomized scheduling, and synchronization mechanisms like Flync, which leverages shared environmental cues \cite{b12}. These approaches are crucial for bootstrapping communication and enabling persistent operation in ultra-low-power scenarios. Furthermore, the complex domain of Wireless Information and Power Transfer (WIPT), especially in multi-user SWIPT contexts, reveals that resource allocation strategies—including beamforming, power control, and interference management—are fundamentally reshaped by the characteristics of distributed energy harvesters, with nonlinear models demanding a significant departure from traditional linear assumptions \cite{b13}.

Across these varied applications, several unifying themes emerge:
\begin{itemize}
    \item The necessity of \textbf{coordination and information sharing} among distributed entities to achieve global system objectives.
    \item The critical importance of \textbf{adaptivity} in resource allocation to cope with dynamic channel conditions, energy availability, and network topology.
    \item The power of \textbf{diversity} (spatial, frequency, temporal, and energy source) when exploited by distributed systems.
    \item The profound impact of \textbf{realistic physical layer modeling}, especially of energy harvesters and intermittent power sources, on the design and efficacy of resource allocation algorithms.
    \item The growing potential of \textbf{repurposing commodity infrastructure} for advanced communication and sensing tasks, thereby lowering deployment barriers.
\end{itemize}

Looking ahead, the continued evolution of IoT and low-power wireless technologies will further accentuate the need for robust, scalable, and efficient distributed systems. Future research will likely focus on developing more autonomous and intelligent resource allocation algorithms, potentially leveraging machine learning and artificial intelligence, to manage the increasing complexity and dynamism of these networks. The pursuit of seamless integration between information transfer, energy harvesting, sensing, and localization within unified distributed frameworks remains a compelling vision. The insights gathered from the reviewed works provide a strong foundation for these future endeavors, paving the way for a truly pervasive and intelligent connected world.

\section*{Acknowledgment}

The preferred spelling of the word ``acknowledgment'' in America is without 
an ``e'' after the ``g''. Avoid the stilted expression ``one of us (R. B. 
G.) thanks $\ldots$''. Instead, try ``R. B. G. thanks$\ldots$''. Put sponsor 
acknowledgments in the unnumbered footnote on the first page.






\vspace{12pt}

\end{document}